# Implications of Stahl's Theorems to Holomorphic Embedding Pt. 1: Theoretical Convergence

Songyan Li, *Student Member, IEEE*, Daniel Tylavsky, *Life Senior Member, IEEE*
Di Shi, *Senior Member, IEEE,* Zhiwei Wang, *Senior Member, IEEE*

*Abstract*—What has become known as Stahl's Theorem in power engineering circles has been used to justify a convergence guarantee of the Holomorphic Embedding Method (HEM) as it applies to the power-flow (PF) problem. In this two-part paper, we examine in more detail the implications of Stahl's theorems to both theoretical and numerical convergence for a wider range of problems to which these theorems are now being applied. In Pt. 1, we introduce the theorems using the necessary mathematical parlance and then translate the language to show its implications to convergence of nonlinear problems in general and the PF problem specifically. We show that, among other possibilities, the existence of Chebotarev points, which are embedding specific, are a possible theoretical impediment to convergence. Numerical impediments to convergence are discussed in the companion paper.

## I. Mathematical Notation

$\mathbb{C}$–Complex plane excluding infinity
$\overline{\mathbb{C}}$–Extended complex plane (includes infinity)
$\text{Cap}(E)$—Capacity of set $E$
$D$—Domain
$D_f$—Convergence domain
$E$—Set containing the singularities of a function
$\mathbb{N}$—Set of natural numbers
$\partial D$—Boundary of domain D

## II. Introduction

Since its introduction to the power-system community in 2012, the holomorphic embedding method (HEM) has attracted a great deal of interest [1], [2]. In addition to being extensively studied as a solution to the power-flow problem, [3]-[7], it has been applied to unit commitment, [8], special devices, [9], [10], reduced-order network equivalents, [11]-[12], estimation of voltage stability margin, [13]-[18], expanded to handle multi-dimensional embeddings [19]-[21], applied to dynamic simulations, [22], used to attack the multivalued problem [23]-[26], and enhanced to improve robustness and computational efficiency, [27]-[30]. One of the reasons for the broad interest in HEM, was the claim of a universal convergence guarantee [1] supported by Stahl's theorems, [34], [35]. While the claim that "any close-to-diagonal sequence of Padé approximants converge in capacity to said function in the extremal domain" [1] is correct (as we will see), published incorrect claims remain, such as, "…if the Padé approximants do not converge at s = 1 …then it is guaranteed that there is no solution (that is, the system is beyond voltage collapse," and "therefore the solution is obtained when it exists, or a divergence is obtained when it does not exist."

The claim of guaranteed convergence was generally (though not universally) accepted by the community, leading partially to the wide-ranging interest. Given that the team at ASU was guilty of too easily accepting the aforementioned claims with insufficient evidence (and—in hindsight—evidence to the contrary at times), it felt incumbent upon the team to look more carefully at these claims. The objective of this paper and its companion paper is to do just that.

The first evidence that all was not as claimed, was that the implementation of the HEM variant as patented in [2] (labeled the Holomorphic Load-Flow Method (HELM)) exhibited convergence problems. The team diagnosed the convergence problem as a numerical problem caused by the Padé approximant (PA) procedure (Viskovatov method) cited in the patent and improved performance was obtained by replacing it with the so-called matrix method, while also changing the embedding [6]. Further, a survey of existing numerical methods commonly used for accelerating the convergence of numerical series showed that, while theoretical convergence may or may not be guaranteed at times, numerical convergence is certainly not guaranteed; further of the numerical methods available, the matrix and Eta methods (both Padé approximant methods) were experimentally demonstrated to be superior to the Viskovatov method, as were other techniques on the systems studied [31].

Despite those numerical successes, convergence difficulties were still encountered; most were resolved one way or another, though some vexing problems remained. Because the conclusions of so many research papers over the broad range of subjects cited earlier were predicated on the claims of universal convergence, it seemed important to us to look closely at what Stahl's theorem did and did not say. At about the time we decided to look more deeply into the implications of Stahl's theorems, [3] also indicated that universal convergence could not be guaranteed. Since then, other researchers have joined in the fray, attempting to address the convergence issues [32].

The present two-part paper looks more closely at the theoretical and numerical convergence underpinnings of HEM. Part 1 of this work makes several contributions. It describes the theorems necessary for understanding the source of the putative convergence guarantee and translates their implications to the range of nonlinear power system problems cited. It shows that any convergence guarantees are embedding specific and shows what must be proven to establish a theoretical convergence guarantee. It shows the centrality of the branch cut to understanding HEM and, in short, that theoretical convergence is dictated by the topology of the branch

Songyan Li and Daniel Tylavsky are with the School of Electrical Computer and Energy Engineering in Arizona State University, Tempe, AZ 85287, USA (email: songyan.l@asu.edu / tylavsky@asu.edu). Di Shi and Zhiwei Wang are with GEIRI-North America, San Clara CA, 95134, USA (email: di.shi@geirina.net / zhiwei.wang@geirina.net ).

cut and numerical convergence is limited by the capacity of branch cut. Based on both the implication of these theorems and the properties of the PAs, predictions are made that help explain the numerical convergence behavior observed in the companion paper.

## III. HEM REVIEW

The HEM method is based on a four-step approach.
1. Holomorphically embed the nonlinear equations appropriately.
2. Represent each embedded variable as their Maclaurin series (germ) and generate a recursion relationship for the Maclaurin series coefficients.
3. The zeroth-ordered (top-of-the-germ/reference state) series coefficient is calculated for each variable and then, using the recursion relationships, the series coefficients are calculated until the desired accuracy is obtained.
4. Padé approximants (PAs) are used to both accelerate and force convergence of the series provided a solution exists in the extremal domain [37], [38].

### A. PF Problem Embeddings and Maclaurin Series Generation

There are an infinite number of ways that nonlinear equations may be embedded. Two common embeddings and two variations are discussed in the companion paper using the nonlinear PF problem as an exemplar. For any given embedding, the complex-valued Maclaurin series generated by HEM are made convergent within the extremal domain (to be defined) by PAs.

### B. Padé Approximants

There is no guarantee that the Maclaurin series generated by HEM will converge at the solution point. Stahl's theorem [34] (examined in some detail later) states, in short, that under some reasonable assumptions, the so-called near-diagonal PAs (discussed later) are the maximal analytic continuation of the function described by the series HEM produces. Given a series with $L+M+1$ terms, the $[L/M]$ Padé approximant is given by:

$$[L/M] = \frac{a[0] + a[1]\alpha + \cdots a[L]\alpha^L}{b[0] + b[1]\alpha + \cdots b[M]\alpha^M} \quad (1)$$

The precise definition of what constitutes a near-diagonal PA is described later, for now we take as a working definition the requirement that the difference, $L-M$, be small. For power flow applications the best selection we have found is $M=L+1$, and we refer to this as an $[M/M+1]$.

## IV. BRANCH POINTS AND BRANCH CUTS

The implications of Stahl's theorems are that theoretical convergence and the convergence rates of HEM (as applied to the PF problem) are determined by the position of the branch points and topology of the branch cuts. In this section we level-set the discussion by establishing a notation and providing some definitions that are central to the discussion.

Consider the complex variable, $\alpha = r\, e^{j\theta}$. Observe that there exists ambiguity in defining $\theta$, all of whose values correspond to the same point in the complex plane. Hence, $g(\alpha) = \ln(\alpha)$ has an infinity of values, $g_k(\alpha)$,

$$g_0(\alpha) = \ln(\alpha) = \ln(r\, e^{j\theta}) = \ln(r) + j\theta,$$
$$0 \leq \theta < 2\pi \quad (2)$$
$$g_k(\alpha) = \ln(r) + j(\theta + k2\pi)$$

### A. Branch Point

A branch point of a multi-valued function in the complex plane is a point where the function exhibits a discontinuity when traversing an arbitrarily small closed circuit which encloses the point. For our example, the essential singularity of $g(\alpha)$ at zero represents a (logarithmic) branch point. (The essential singularity in the PF problem is the saddle-node bifurcation point (SNBP), which is an algebraic branch point.)

### B. Branch Cut

For the inverse of a function to exist, and to avoid ambiguity, the mapping from argument to function value must be one-to-one or, equivalently, injective. Avoiding the discussion of Reimann surfaces, we may eliminate the ambiguity by cutting the complex plane with a branch cut that restricts our $\ln(\cdot)$ function to be in the any range covering $2\pi$ radians. (This is similar to the usual limiting of the $\arcsin(\cdot)$ function to the interval $[-\pi/2, \pi/2]$.) There are infinitely many branch cuts which accomplish our goal; a common one is the ray extending from the origin to positive infinity along the real axis, (Fig. 1).

While the selection of the branch cut in this case is somewhat arbitrary, we will see that a branch cut with a specific property, that of minimum logarithmic capacity, will exhibit properties that impact convergence behavior of HEM.

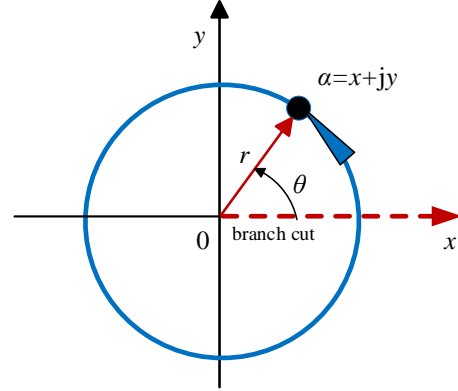

Fig. 1 Branch cut and branch point definition

## V. STAHL'S THEOREMS

While we will make this more precise, the essence of Stahl's theorems has been interpreted to say that the use of PAs guarantees convergence of the PF problem to any loading short of the SNBP. While this is theoretically likely for most PF problems using one of the embeddings shown in the companion paper, we show in this paper that this is not necessarily the case. (In the companion paper we also show that the numerical convergence domain may differ from the theoretical convergence domain.)

### A. Stahl's Assumptions

This section contains a summary of the assumptions and most important definitions needed to understand the implications of Stahl's theorem (Theorem 1.4 [34]) for functions with branch points. (For functions without branch points see [36].)

Assume a function, holomorphic in $\alpha$ in a neighborhood at infinity, with an expansion given by,

$$f(\alpha) = f[0] + \frac{1}{\alpha}f[1] + \frac{1}{\alpha^2}f[2] + \cdots \quad (3)$$

is to be approximated using a PA.

*Remarks* (1) We will eventually see why it is important to expand about infinity. For the power system (PS) problems, nonexistence of a solution at infinity will require expansion about zero. Since all of the convergence theorems use (3) as their basis, we will stay with this notation and show the changes that need to be made when we apply these theorems to the PS-type problems.

(2) We will refer to the plane in which the roots of the PA of the expansion in (3) are plotted as the 'inverse α plane' and the plane containing the roots of the PAs of functions expanded about zero as the 'α plane.'

Stahl's fundamental question is: *Over what domain will the approximation of* (3) *by PAs be valid?*

In Stahl's first group of results, the following assumptions are made.

*Assumptions* 1.1 [34]: The function, $f$, is (i) assumed to be analytic at infinity and (ii) to have all of its singularities in a compact set $E \subseteq \overline{\mathbb{C}}$ with $cap(E) = 0$, i.e., $f$ has analytical continuations along any path in $\overline{\mathbb{C}} \backslash E$ starting at infinity; the continuation may be multiple-valued.

*Remarks*. (1) The first step, which is the most difficult and whose development here is specific to the PF problem (i.e., polynomial equations), is to show that the *solutions* to the PF equations obey these assumptions. The arguments made toward this goal in [1] are correct but incomplete and do not demonstrate how those arguments are connected with Stahl's assumptions. We remedy that minor oversight below, providing the argument that the all bus voltages, $V(\alpha)$, obtained using the classical embedding in the companion paper are holomorphic functions except at a set of discrete exceptional points and connecting this argument with Stahl's assumptions.

Without loss of generality, we limit this development to PQ buses and rewrite the bus power equilibrium equations, (4), as two equations, (5), to eliminate the complex conjugate operator,

$$V_i^*(\alpha^*) \sum_{k=1}^{N} Y_{ik} V_k(\alpha) = \alpha S_i^* \tag{4}$$

$$\overline{V}_i(\alpha) \sum_{k=1}^{N} Y_{ik} V_k(\alpha) = \alpha S_i^* \tag{5}$$
$$V_i(\alpha) \sum_{k=1}^{N} Y_{ik}^* \overline{V}_k(\alpha) = \alpha S_i$$

where $V_i(\alpha)$ and $\overline{V}_i(\alpha)$ are independent bus $i$ voltage functions, $S_i$ is the complex power injected into bus $i$, and $Y_{ik}$ is an entry of the bus admittance matrix. Only those solutions of (5) that are consistent with the reflection condition, (6), are of interest.

$$\overline{V}_i(\alpha) = V_i^*(\alpha^*) \tag{6}$$

Solutions that do not obey this reflection condition are non-physical solutions to the PF problem. Given that (5) represents a set of complex polynomial equations, the theory of resultants and Gröbner basis can be used through a triangularization-like process of elimination to produce a polynomial in $V_1(\alpha)$, and polynomial in α, $p_n(\alpha)$, provided it exists,

$$\mathfrak{P}(\alpha, V_1) = \sum_{n=0}^{N} p_n(\alpha) V_1^n(\alpha) = 0 \tag{7}$$

and by a simple 'back substitution' process, all voltages can be obtained similarly. While the degree of the polynomial in (7) is typically of exponential order in the number of variables in the original equation, $N$, the order is still finite. The roots of (7) represent all of the solutions of (5) (including those that do not conform to (6)) as functions of α. The complex-valued function, $V_1(\alpha)$, is defined implicitly in (7), which is the definition of an algebraic curve and, therefore $V_1(\alpha)$ is holomorphic for values of α where the conditions of the implicit function theorem are satisfied [1]. Two methods in the next paragraph can be used to test whether Stahl's assumption (i) is satisfied for PF problem specifically. To show that condition (ii) of Assumption 1.1 is satisfied, recognize that, by the implicit function theorem, $V_1(\alpha)$ is holomorphic in the neighborhood of all values of α, except at points where,

$$\frac{\partial \mathfrak{P}(\alpha, V_1)}{\partial V_1} = 0 \tag{8}$$

The singularities of $V_1$ are found by simultaneous solving polynomials (7) and (8). Recognizing that since $\mathfrak{P}$ is a finite degree polynomial in $V_1$, and since $p_n$ is a finite degree polynomial, the Gröbner basis approach applied to the solution of (7) and (8) must yield another finite degree polynomial in $V_1$, $\mathfrak{F}(V_1) = 0$. Because $\mathfrak{F}(V_1)$ is a finite degree polynomial, the solutions to $\mathfrak{F}(V_1) = 0$ must be a set of discrete points whose cardinality must be bounded, and hence of capacity zero. This satisfies criterion (ii).

For HEM to be useful when applied to the PF problem, $V(\alpha)$ must be holomorphic at $\alpha = 0$. Two ways to address this issue are: By the implicit function theorem, if the Jacobian of the problem, (5), is nonsingular at a point, then the implicitly defined functions are holomorphic at that point. An alternative practical way to address this issue, is to observe the Maclaurin series generated by the HEM formulation of interest and, if the series has a non-zero radius of convergence (ROC) (around $\alpha = 0$), then $V(\alpha)$ is by definition analytic and (from a major theorem in complex analysis [33]) complex-valued analytic functions are holomorphic. (Recall that a non-zero ROC of $V$ guarantees that its complex derivative exists everywhere in a neighborhood at the origin.) The most practical way is to see if the HEM based algorithm converges or, alternatively, converges in a neighborhood of $\alpha = 0$. Given that $\alpha = 0$ typically corresponds to either a no-load or lightly loaded condition, the voltage functions are typically well behaved and holomorphic for these cases. While this argument has been made for the PF problem, its extension to the range of PS problems is straight forward.

It is worth taking a moment to understand the implications of (7). For any given α, the roots of (7) correspond to physical solutions (those that obey (6)), and non-physical solutions (those that do not obey (6)). Of the physical solutions, there is debate about what constitutes an 'operable' solution. Let's set that debate aside as outside the scope of this paper and agree that there is a least one root of (7) for each value of the embedding parameter, α, that we would consider an operable solution…even if we do not all agree on the same root. Ideally, the operable solution would be continuous, not encountering a branch point as α is varied from no-load to the SNBP; however, the range over which this solution is continuous is a function of the embedding. It is possible to construct embeddings that cause the solution of the embedded problem to disappear short of the SNBP (has a branch point) even though the PF problem has a well-defined solution throughout the range. The convergence domain (discussed next) of Stahl's theorems factors in the existence of premature branch points created by a poor choice of embedding. (How to embed a problem so that premature branch (and Chebotarev) points are guaranteed not to occur is an active area of research and beyond the scope of this work.

### B. Stahl's Convergence Domain Existence and Uniqueness Theorem

The objective of the proof of the first of Stahl's theorems is to show that a convergence domain for PAs exists and is unique.

Theorem 1.1 [34] (Abbreviated) *Let f satisfy Assumptions 1.1. Then there exists a domain $D = D_f \subseteq \bar{\mathbb{C}}$, $\infty \in D$, which is unique up to a set of capacity zero, and*

(i) *the sequence $\{[m_j/n_j]\}_{j \in \mathbb{N}}$ of Padé approximants converges in capacity to f in the domain D for any sequence $\{(m_j, n_j)\}_{j \in \mathbb{N}}$ of indices satisfying*

$$m_j + n_j \to \infty, \qquad \frac{m_j}{n_j} \to 1 \text{ as } j \to \infty \qquad (9)$$

*Remarks.* (1) This theorem establishes the uniqueness of the convergence domain (unique up to a set containing individual points, but no continuum) for the sequence of near-diagonal PAs, defined by (9).

(2) Convergence in capacity means that we cannot guarantee that all points on the curve will converge (*cf.* pointwise or uniform convergence); we can only guarantee convergence in capacity (defined below). This effectively means that as $m_j + n_j \to \infty$, the capacity of the set where the difference between the function and its approximation is greater than a threshold tends toward zero. Said another way: with an infinite number of terms in the series, the PA is within tolerance everywhere except at (possibly) a countable [36] set of isolated points. More formally:

Defn: Convergence in Capacity: The sequence of Padé approximants, $f_n$, n=1, 2, ..., is said to converge in capacity to $f$ in the domain $D_f \subseteq \bar{\mathbb{C}}$ if for every $\varepsilon > 0$ and every compact set $V \subseteq D \setminus \{\infty\}$ we have $cap\{z \in V | |f(\alpha) - f_n(\alpha)| > \varepsilon\} \to 0$ as $n \to \infty$.

Given that all implementations of HEM use a finite number of terms and finite precision, voltage functions not converging in capacity for short continua have been seen by the authors. But these continua occur near the SNBP for reasons discussed in the companion paper. *Nonconvergence for these continua is a numerical issue, not a theoretical issue*.

### C. Stahl's Convergence in Capacity and Rate Theorem [34]

Theorem 1.2. (Not reproduced here.)

*Remarks.* (1) This theorem defines convergence in capacity as it is to be applied in the context of the development these theorems. We shall not need the specifics of this definition.

(2) This theorem also establishes the convergence rate within the convergence domain as linear and gives the asymptotic convergence *factor* as $\alpha \to \infty$. Note that a convergence factor less than (equal to) 1 indicates convergence (nonconvergence). The asymptotic convergence rate inside the convergence domain, $G_D(\alpha)$, is given by,

$$G_D(\alpha) = \frac{cap(\partial D)}{|\alpha|} + O(\alpha^{-2}) \text{ as } \alpha \to \infty \qquad (10)$$

Eq. (10) implies that the convergence *rate*, $G_D^{-1}(\alpha)$, is greatest near infinity for an expansion about infinity, (3), or, equivalently, greatest near the point of expansion. (As shown in the companion paper, this prediction is consistent with numerical simulations where, for the PF problem, (10) is adjusted for an expansion about zero.) It is well known that the convergence rate of Taylor series is highest near the point of expansion and decreases as we move away from this point, so it is not surprising that PAs, constructed from the Taylor polynomials, should have similar behavior. When a function has no branch point, $G_D(\alpha)$ is 0 everywhere [34], which implies that the appearance branch points affects the convergence rate.

### D. Stahl's Uniqueness of the Extremal Domain Theorem

Theorem 1.3 [34] ([37], [38] Theorems 1 and 2). *Let the function f be analytic at infinity. Then there uniquely exists a domain $D \subseteq \bar{\mathbb{C}}$ satisfying the following three conditions:*

(i) $\infty \in D$ *and the function f has a single-valued analytic continuation in D.*

(ii) $cap(\partial D) = inf_{\tilde{D}} cap(\partial \tilde{D})$, *where the infimum extends over all domains $\tilde{D} \subseteq \bar{\mathbb{C}}$ satisfying assertion* (i)

(iii) $D = \bigcup \tilde{D}$, *where the union extends over all domains, $\tilde{D} \subseteq \bar{\mathbb{C}}$ satisfying the assertions* (i) *and* (ii).

Remarks. (1) The domain, *D*, defined by this theorem is unique and is referred to as the *extremal* domain for the single-valued analytic continuation of the function *f*.

(2) Given that a PA is single valued, it is not surprising that assertion (i) is required.

(3) Without assertion (iii) uniqueness is determined only up to a set of zero capacity. This point is minor but insightful in two ways. Since the extremal domain as defined in (ii) is (effectively) an optimization problem and because isolated points have capacity zero, the capacity of the continuum that is the boundary of *D*, $\partial D$ (the branch cut and possibly some other isolated points), is unchanged by adding to $\partial D$ a set of isolated points; hence the minimization problem has an infinite number of solutions. The extremal domain *D* is the union of this infinite number of solutions, which eliminates any isolated points not on $\partial D$.

(4) Note that function *f* may have isolated poles but these poles do not affect the uniqueness described in (3) because a pole (i.e., isolated singularity) is not part of the convergence domain. At the function's pole, convergence cannot be defined, and the function is not defined. While, the PA will still converge around the isolated pole, it will never converge on the branch cut with minimum logarithmic capacity. This point will be demonstrated in a future paper.

(5) Assertion (ii) defines the requirement for the boundary of the extremal domain. This assertion does not say how to calculate this domain, but a formal statement of the minimization problem may be found in [36]. What we are able to prove, and is the subject of a forthcoming paper, is that *the (ultimate) root distribution on the branch cut defined by the PA is identical to the electrostatic charge distribution at equilibrium on a conductor with the two-dimensional topology of the branch cut as it exists in the inverse $\alpha$ plane*. In the companion paper we will see that a poor embedding can lead to numerical non-convergence, despite the theoretical guarantee of convergence.

*E. Stahl's Convergence Theorem*

The following theorem based on Assumptions 1.1 (or Stahl's related Thm. 1.7 [34] based not on Assumption 1.1 but on a symmetry property) is what authors in the power system literature refer to as Stahl's theorem.

Theorem 1.4 [34] *If the function f satisfies Assumption 1.1, then the convergence domain, $D_f$,* (of the sequence of close-to-diagonal Padé approximants) *of Theorem 1.1 is identical with the extremal domain D of Theorem 1.3, up to a set of capacity zero.*

Remarks. (1) The theorem is fairly straightforward. The convergence domain of the sequence of near-diagonal PAs defined in Theorem 1.1 is identical with the extremal domain calculated using the assertion (ii) of theorem 1.3, up to a set of isolate points (a set of capacity zero).

(2) The boundary of $D$, $\partial D$, is the location where the poles (and zeros) of the Padé approximant tend to accumulate since this is (by theorem) the location where the PAs do not converge; however some (spurious) poles of the PA may cluster inside $D_f$, leading to places where the PAs will (by definition) not converge. (Nuttal's conjecture is that the number of elements in this set is bounded [39] and therefore has capacity zero. The precise location of these poles changes as the $m_j + n_j$ changes.)

(3) The proof of this theorem shows that the extremal domain is at most a subset of the convergence domain, $D \subseteq D_f$, and the difference, $D_f \setminus D$, is a set of capacity zero.

(4) Our experience with solving the PF problem is that we often see spurious poles inside $D_f$, as discussed in the companion paper and (rarely) some very close to (but not on) the real line, sometimes at the origin (often in the form of Froissart doublets), but have not experienced any nonconvergence issues due to these poles, except of course at the pole itself. So, if the pole occurs on the real line at/near the point of interest, convergence can be affected.

(5) When discussing Stahl's work, maximal is shorthand used to mean (essentially) that the boundary of the extremal domain, $\partial D$, whose continua are synonymous with the branch cut, is of minimal (logarithmic) capacity among all compact sets that could be used to define an appropriate branch-cut for a single-valued *f*. Maximal analytic continuation refers to the ability of the near-diagonal PAs to analytically continue the series representing the function beyond its region of convergence to the entire convergence domain of the function, absent a set of poles inside $D_f$.

(6) Any confusion over whether or not Stahl's theorem conveys a *theoretical* convergence guarantee is caused by the subtle difference between the function's domain, the extremal domain, and the domain we are interested in, the convergence domain, which is the domain in which the sequence of near-diagonal PAs theoretically converge (in capacity) to the function. The convergence domain is identical to the extremal domain (up to at a set of isolate points) but is not identical with the function's domain. It is because the convergence domain and the function's domain are not identical that theoretical non-convergence can be encountered. In the next section we make this distinction clear with examples and show one source of theoretical nonconvergence.

## VI. Convergence Domain, Extremal Domain, Function's Domain, Chebotarev Points and Branch Cuts

One of the properties of PAs is that, while they can continue a function through a simple pole, they cannot continue a function through a branch point or (what is known as) a Chebotarev point. It is easier to introduce these concepts using logarithmic functions, which we now do, and then extend them to the PF problem in the companion paper.

Consider the logarithmic functions in (11), where the branch points of the functions for cases A-D are given in Table 1.

$$h(z) = \begin{cases} \ln\left[\dfrac{(z-a_1)(z-a_3)}{(z-a_2)(z-a_4)}\right] (A,B) \\ \ln\left[\dfrac{(z-a_1)(z-a_3)(z-a_5)}{(z-a_2)(z-a_4)(z-a_6)}\right] (C,D) \end{cases} \quad (11)$$

Table 1 Pole Values for Case A-D

| Root\Case | A | B | C | D |
|---|---|---|---|---|
| a1 | 2+j3 | 2+j1.5 | 2+j1.5 | 2+j1.5 |
| a2 | -2+j3 | -2+j1.5 | -2+j1.5 | -2+j1.5 |
| a3 | 2-j3 | 2-j1.5 | 2-j1.5 | 2-j1.5 |
| a4 | -2-j3 | -2-j1.5 | -2-j1.5 | -2-j1.5 |
| a5 |  |  | -1.6 | -0.7 |
| a6 |  |  | +1.6 | +1.6 |

Consider the near-diagonal [99/100] PA for Cases A-D shown in Fig. 2-Fig. 5 constructed from a series expanded about infinity and plotted in the inverse-α plane. The series and PAs were calculated using high precision to avoid anomalies due to roundoff errors.

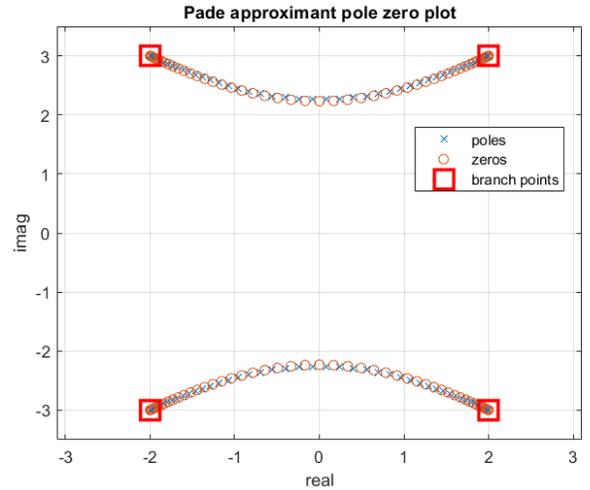

Fig. 2 Branch cut for Case A, inverse α plane.

Remarks: (1) Observe in Fig. 2-Fig. 5 (cases A-D) that: (a) the roots of the PA accumulate on the branch cut, the continua of $\partial D$; (b) the domain of the function, which is the complex plane absent the branch points of the function, i.e. $\overline{\mathbb{C}} \setminus E$, is larger than the convergence domain, which is $D_f$.

(2) For Case A in Fig. 2 the branch cut is made up of the union two disjoint open analytic Jordan arcs, each terminated in two branch points. Note: $h(0)$ for Case A is a point of PA convergence.

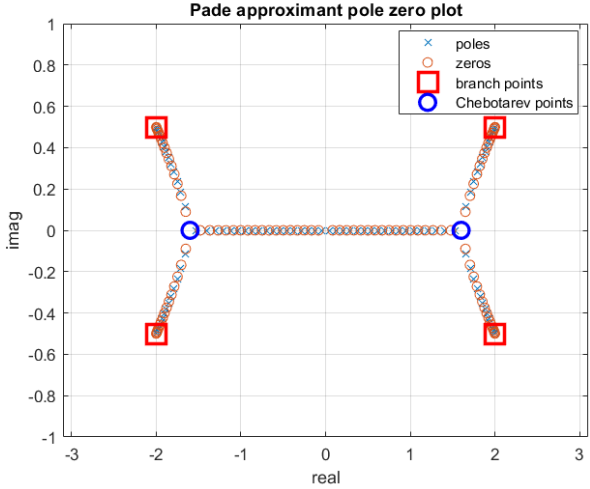

Fig. 3 Branch cut for Case B, inverse $\alpha$ plane.

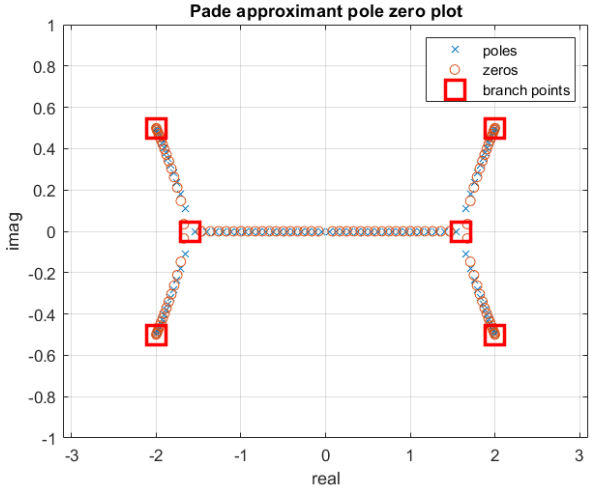

Fig. 4 Branch cut for Case C, inverse $\alpha$ plane.

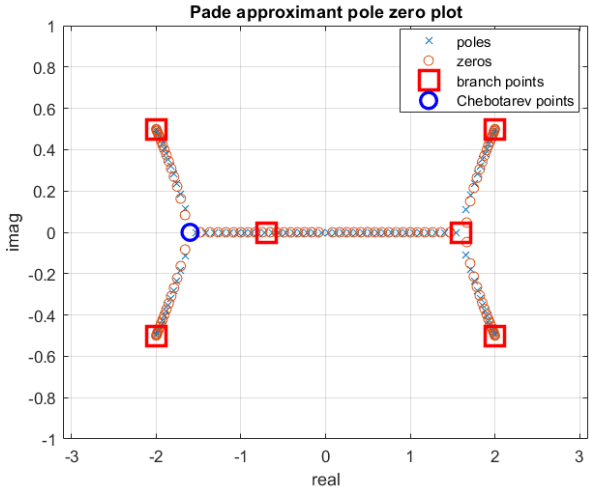

Fig. 5 Branch cut for Case D, inverse $\alpha$ plane.

(3) When the imaginary parts of the branch points are reduced from j3.0 in Case A to j1.5 Case B, the branch cut takes on a different topology (dictated by Thm. 1.3). The points on the real axis at ±1.6 are known as a Chebotarev points: the juncture of the Jordan arcs at other than a branch point. PAs cannot continue a function through a Chebotarev point, since a Chebotarev point is part of the branch cut. The pole density at a Chebotarev point is theoretically zero. Notice that while $h(0)$ is in the function's domain, it is no longer a point of convergence, however (-1.6-$|\varepsilon|$), $\varepsilon \neq 0, \varepsilon \in \mathbb{R}$, is in $D_f$.

(4) If we construct our function by starting with Case B and add additional branch points at the Chebotarev points, we get Case C in Fig. 4, which is identical to Case B, except for the distinction that no Chebotarev points exist and the branch cut is made up of the union of five disjoint open analytic Jordan arcs, each terminated in two branch points.

(5) If we start with Case C and move the one branch point located at -1.6 to -0.7 to get Case D, the topology of the branch cut is unchanged, but our function, developed at infinity, can no longer be continued along the real axis and approach the branch point at -0.7 because of the Chebotarev point at -1.6 in Fig. 5: the boundary of the region of convergence (branch cut) prevents the analytic continuation of the function from -1.6 toward the origin even though that is part of the domain of the function and the function is well defined in that region. A Chebotarev point 'covering' the branch point on the real axis that corresponds to the SNBP is the case of concern in guaranteeing PF convergence. The point of this discussion is this: Presently there is no holomorphic embedding of the PF problem that has been proven to guarantee that a Chebotarev point will not prevent the PF from converging for loadings up to the SNBP. *To prove that convergence is guaranteed to an operable solution that exists, one must prove that, (i) $V(\alpha)$ is holomorphic at the point of development and at the solution point, (ii) $V(\alpha)$ is holomorphic along the real axis except at a set of isolated singularities, (iii) the solution point does not reside on the branch cut with minimum logarithmic capacity and (iv) this branch cut does not intersect the real axis short of the solution point.*

## VII. CHEBOTAREV POINTS AND EXPANSION ABOUT ZERO

We refer to the planes of Fig. 2-Fig. 5 as the inverse $\alpha$ plane because these functions were developed at infinity, consistent with (3). Had Stahl developed these functions at zero, the branch cuts in the '$\alpha$ plane' would be infinite and the capacity of these branch cuts would likewise be infinite, perhaps making proof of the convergence theorems impossible, or at least more difficult.

Observe that we could just as easily have developed the series describing the Cases A-D about either zero or infinity, but for the PS problems, where the functions are undefined at infinity, we must develop the function about zero as shown in (12).

$$f(\alpha) = f[0] + \alpha f[1] + \alpha^2 f[2] + \cdots \quad (12)$$

While, in the companion paper, will work at times with branch cuts for the PF problem in the $\alpha$ plane, we must keep in mind that all of the results and intuition about how the branch cuts should behave must be within the context of the inverse $\alpha$ plane.

In working with functions developed both at infinity and zero, we have noticed a curious effect that will show up for the PF problem. Consider the following function:

$$h(z) = \ln\left[\frac{(z-1)}{(z+1)}\right] \quad (13)$$

If we build the PA for a series developed about zero, the notation used is that shown in (14). (For functions developed about infinity, this is indicated as $[L/M]_{\infty, \frac{1}{\alpha}}$.)

$$[L/M]_{0,\alpha} = \frac{a[0] + a[1]\alpha + \cdots a[L]\alpha^L}{b[0] + b[1]\alpha + \cdots b[M]\alpha^M} \qquad (14)$$

The roots of the PA in (14) in the inverse α plane are the roots of:

$$[L/M]_{0,\frac{1}{\alpha}} = \frac{a[0] + a[1]\frac{1}{\alpha} + \cdots a[L]\frac{1}{\alpha^L}}{b[0] + b[1]\frac{1}{\alpha} + \cdots b[M]\frac{1}{\alpha^M}} \qquad (15)$$

The roots of the $[25/26]_{\infty,\cdot}$ PA for $h(z)$ in (13), developed at infinity, in the inverse α plane and the α plane are shown in Fig. 6 and Fig. 7, respectively. Observe that the branch cut falls along the real axis.

Compare Fig. 6 and Fig. 7 to Fig. 8 and Fig. 9, respectively, where (13) was developed about zero and a $[25/26]_0$ PA produced. Note that the roots of the PA are no longer aligned with the real axis in Fig. 8 in the inverse-α plane, and this effect is made more evident in Fig. 9 when plotted in the α plane. (Had we chosen a larger plot range, we would have seen that both the poles and zeros in Fig. 9 form closed arcs.) As we add more terms in the series expansion, the poles and zeros converge toward the real axis as shown in Fig. 10 in the inverse α plane (and the in the alpha plane not shown). Why this difference exists remains an open question. This behavioral difference is introduced here because it affects our interpretation of the plots developed for the PF problem in the companion paper.

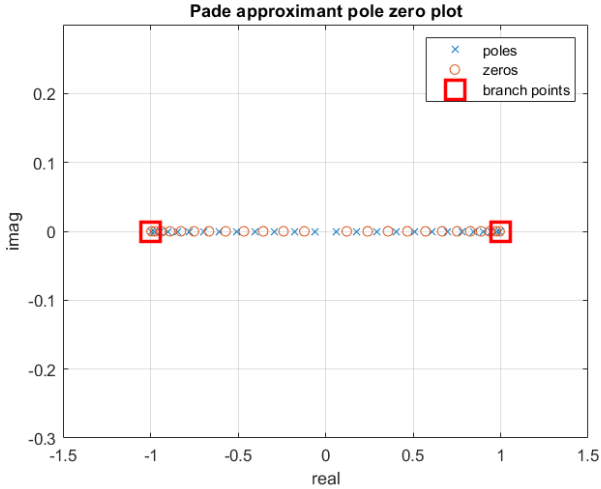

Fig. 6 Roots of $[25/26]_{\infty,\frac{1}{\alpha}}$ (inverse α plane) for (13)

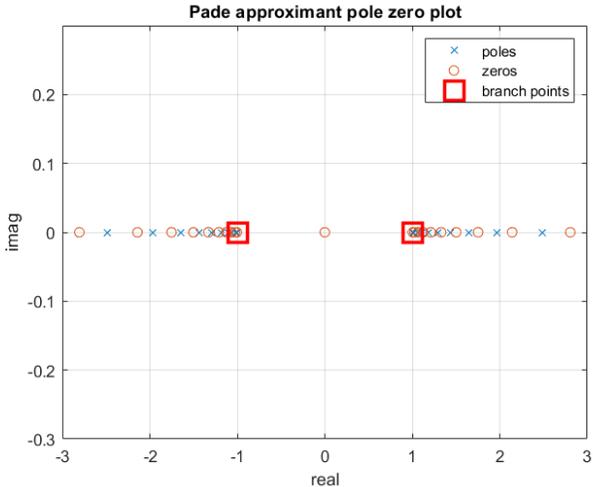

Fig. 7 Roots of $[25/26]_{\infty,\alpha}$ (α plane) for (13)

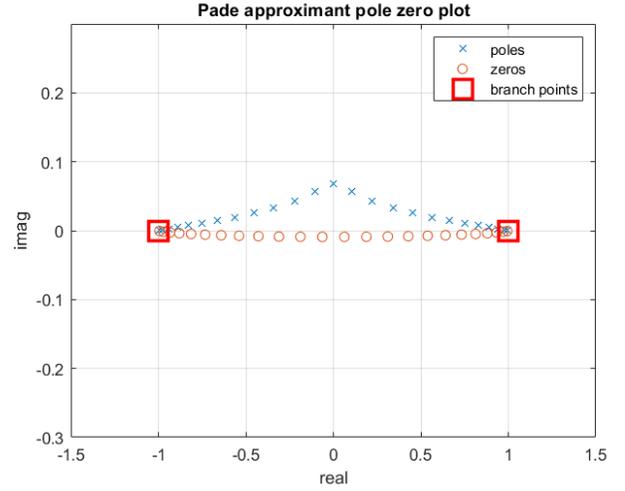

Fig. 8 Roots of $[[25/26]]_{0,\frac{1}{\alpha}}$ (inverse α plane) for (13)

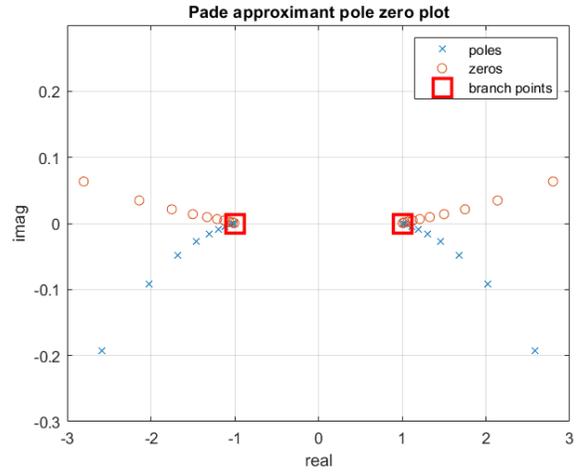

Fig. 9 Roots of $[25/26]_{0,\alpha}$ (α plane) for (13)

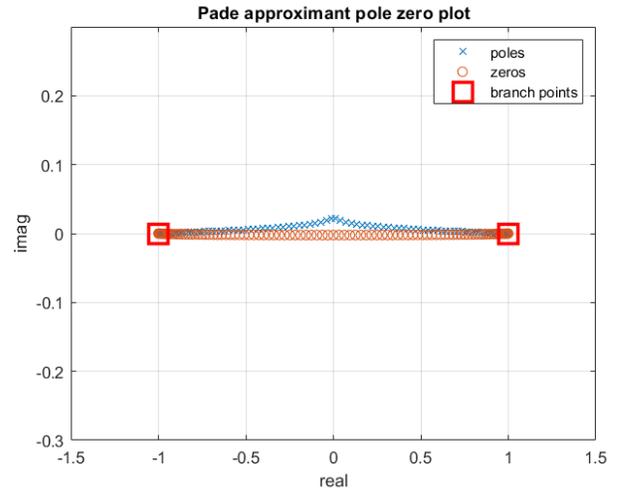

Fig. 10 Roots of $[99/100]_{0,\frac{1}{\alpha}}$ (inverse α plane) for (13)

## VIII. Conclusion

The goal of this paper was to clarify the theoretical implications of Stahl's theorem to the putative convergence guarantee of the PF problem. We show that convergence guarantees are embedding specific and give the conditions that must be satisfied to provide a theoretical (though not numerical) convergence guarantee for any embedding. We have shown that convergence in capacity, rather than pointwise or uniform convergence, is linear and, for series that are less than infinite in length, continua in the function's convergence domain could exist where the PAs are not numerically convergent. In short, theoretical convergence is dictated by the topology of the branch cut and (shown in Part II) numerical convergence is limited by the capacity of branch cut. We state, with a proof in a future publication, that the (ultimate) root density of the PAs on the branch cut is the same as the equilibrium distribution of electrostatic charge on a conductor in 2-D space whose geometry is the same as the branch cut's topology in the inverse α plane.

ACKNOWLEDGEMENT

The authors gratefully acknowledge the support for this work provided through the Power System Engineering Research Center (PSERC) by SGCC Science and Technology Program under contract no. 5455HJ160007.